\documentclass[%
 aip,
 sd,%
 amsmath,amssymb,
 reprint,%
]{revtex4-1}

\usepackage{graphicx}
\usepackage{dcolumn}
\usepackage{bm}

\begin{document}


\title{Planar graphene tunnel field-effect transistor} 



\author{V. L. Katkov}
\email[]{katkov@theor.jinr.ru}
\author{V. A. Osipov}
\affiliation{Bogoliubov Laboratory of Theoretical Physics, Joint
Institute for Nuclear Research, 141980 Dubna, Moscow region,
Russia}


\date{\today}

\begin{abstract}
  We propose a concept for a graphene tunnel field-effect transistor. The main idea is based on the use of two graphene electrodes with zigzag termination divided by a narrow gap under the influence of the common gate. Our analysis shows that such device will have a pronounced switching effect at low gate voltage and high on/off current ratio at room temperature.  
\end{abstract}


\maketitle 

Since its discovery in 2004, graphene has been considered as a promising material for nanoelectronics which could replace silicon technologies. However, a significant problem that makes it difficult to use graphene in transistors and digital circuits is the absence of band gap and, therefore, impossibility to switch off graphene. Many attempts have been done to find ways of creating an artificial band gap using various methods such as applying electric fields, doping with atoms, stretching and squeezing the material, etc. These approaches have made possible to produce band gaps in the few hundred meV while practical digital circuits require a band gap on the order of 1 eV at room temperature.

Recently, a serious breakthrough has been done in the fabrication of atomically smooth graphene edges \cite{Kim,GE1,GE2,GE3,GE4,NNG1}. This stimulates an intensive study of graphene as a material for electrodes in various nanoelectronic devices  \cite{Electrodes1, Electrodes2}. In particular, effects of edge states in plane junctions with graphene electrodes, which are formed by a single-level system  placed between the edges of two single-layer graphene half planes have been investigated \cite{Rk}. It should be noticed that both electrode-molecule-electrode and electrode-electrode devices are of interest in modern electronics. The planar structure of graphene offers promising possibilities for fabrication of tunnel field-effect transistor (TFET) \cite{TFET1}. 
It has been suggested to use either applied electric field  \cite{TFETG1,TFETG2,TFETG3} or an insulator layer  \cite{Fiori,Svintsov} as a tunnel barrier between graphene electrodes. An exciting example of the TFET based on vertical graphene heterostructures has been reported \cite{Geim1, Geim2}, where quantum tunneling from a graphene electrode through a thin insulating barrier (hexagonal boron nitride or molybdenum) is explored. The operation of the device relies on the voltage tunability of the tunneling density of states (DOS) in graphene and of the effective height of the tunnel barrier adjacent to the graphene electrode. Theoretical study of a similar structure has been performed by using the Bardeen transfer Hamiltonian approach \cite{Feenstra}. Notice also that 2D-2D tunneling has been experimentally carried out on coupled electron gas systems in closely placed quantum wells in AlGaAs/GaAs heterostructures \cite{2D1,2D2,2D3,2D4}. 

What is important, a graphene nanogap device with crystallographically matching edges has been recently fabricated \cite{NNG1}. The tunnel barrier here was governed by a graphene nanogap with few hundred nanometers separation in a high-vacuum chamber.  The parallel edges help to build uniform electrical field and allow to perform electron emission study on individual graphene. It was found that emitted electrons are almost independent of the gate voltage within the range from -80 to +80 V. Gateable current-voltage (I-V) characteristics at room temperature were observed in devices fabricated by depositing molecules inside a few-layer graphene nanogap with the gaps having separations of the order of 1-2 nm \cite{Prins}. 

In this letter, we report on a graphene-based-device concept for the TFET. The main idea is based on the use of specific edge state effects in graphene electrodes with zigzag termination in the regime of tunnel current. It has been found that the singular electronic states arise at the Fermi level, whose wave functions are mainly localized on the zigzag edge\cite{NovoselovRev}. This results in remarkably sharp peaks in density of electronic states near the Fermi level. Our calculations show that these peculiar states will markedly modify the tunnel current under the influence of gate voltage. The schematic picture of the device is shown in Fig. 1. One has to satisfy three conditions: (i) both graphene electrodes must be oriented towards each other with crystallographically matching zigzag edges, (ii) in order to reach the regime of tunnel current the gap between contacts must be narrow, and (iii) the gate has influence on both electrodes.

\begin{figure}
\includegraphics[width=6cm]{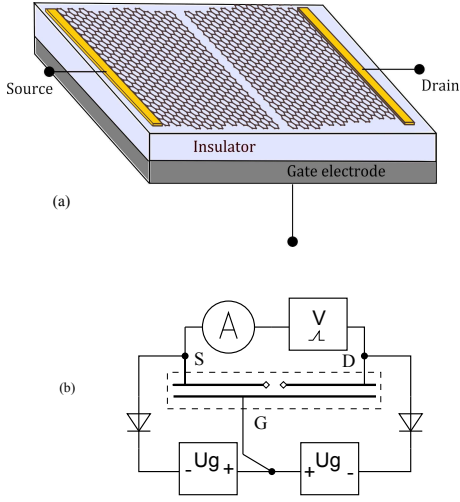}
\caption{Atomic structure of a junction (a) and a possible schematic diagram of the I-V measurement setup (b).}
\end{figure}

\begin{figure}
\includegraphics[width=8cm]{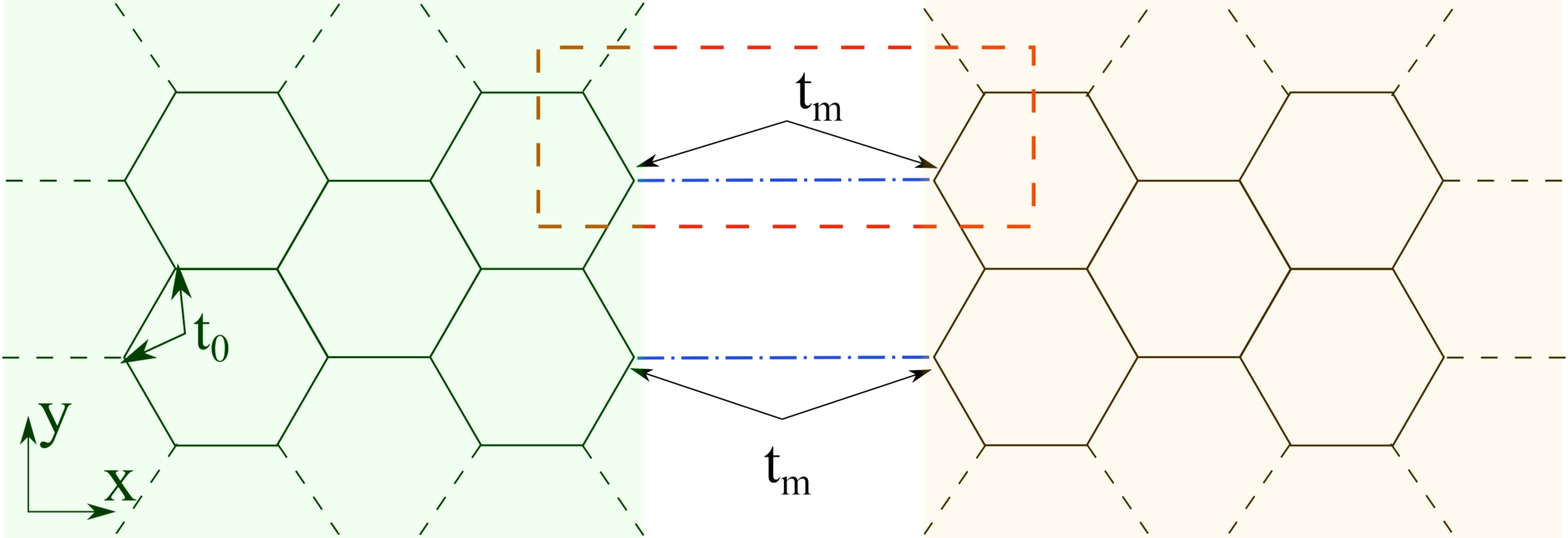}
\caption{Atomic structure of a junction.}
\end{figure}

The atomic structure of graphene contacts is shown in Fig. 2. We assume that electrodes are semi-infinite in the $x$-direction and infinite in the $y$-direction. The calculations are performed using a tight-binding  model combined with the Green's function method. Within this approach the tunnel current is written as \cite{BG, TD}
\begin{eqnarray}\label{GF1.1}
I = &&\frac{e}{h} \int\limits_{-\infty}^{\infty}\mathrm{Tr}[A_L(\mathbf{1}-t^\dag g_R^{-}tg_L^{-} )^{-1}t^{\dag}A_R t(\mathbf{1}-g_L^{+}t^{\dag}g_R^{+}t )^{-1}]\nonumber \\ 
 &&\times [f_L-f_R]d\varepsilon,
\end{eqnarray}

where $g^{+}_{R,L} = [(\varepsilon +i 0^+)\mathbf{1} - H_{R,L}]^{-1}$ is the retarded Green function for the right (left) contact, $g_{R,L}^{-}=(g_{R,L}^{+})^{\dag}$,  $A_{R,L} =i(g_{R,L}^{+}-g_{R,L}^{-})$ is the spectral density. The local density of electronic states for the $n$-th atom is determined as $\rho_l^n = -\mathrm{Im}(g_{nn})/\pi$. The tight-binding Hamiltonian reads
\begin{eqnarray}
H_{L,R} =&& (\pm eV/2+ eV_g)\sum\limits_{i} c_{(L,R),i}^{\dag}c_{(L,R),i}\nonumber\\ 
&&+t_{ij}\sum\limits_{i,j} c_{(L,R),i}^{\dag}c_{(L,R),j},
\end{eqnarray}
where $c_{(L,R),i}^\dag$ $(c_{(L,R),i})$ is the electron creation (annihilaton) operator on $i$th orbital for the left (right) contact, respectively, $V$
is bias voltage, $V_g$ is gate voltage applied to the contacts, 
\begin{equation}
t = \sum\limits_{i,j}t_m^{ij} c_{L,i}^\dag c_{R,j}
\end{equation}
 is an operator describing transitions between the left and the right contacts (see Fig.~2), $t_{ij} = t_0$ and $t_m^{ij} = t_m$ for nearest neighbour atoms and set to be zero for others, $f_{L,R} = f(\varepsilon\pm eV/2)$,  $f(\varepsilon)=[1+\exp(\varepsilon/{k_B T})]^{-1}$ is the Fermi distribution function, $k_B T = 0.01 t_0$ (this value is chosen to be close to the room temperature), $\mathbf{1}$ is the unit matrix. Green functions for edge atoms were calculated within the iteration procedure \cite{Sancho1, Sancho2}. 

The energy and voltage are expressed in terms of $t_0(\approx 2.8 $ eV$)$\cite{NovoselovRev}, the hopping integral of graphene. The interaction parameter $t_m$ is governed by the width of the gap between electrodes. We use the value $t_m=0.01 t_0$, which gives roughly the gap width $d \sim 3a_0$ in accordance with the known relation $t(d) = t_0 \exp[-2(d-a_0)/a_0]$ \cite{HKS, MA}. 
The tunnel current is normalized to the red rectangular cell pictured in Fig. 2.

The results are shown in Fig. 3. As is seen, a pronounced switch-off effect takes place at nonzero gate voltage.  
\begin{figure}
\includegraphics[width=8cm]{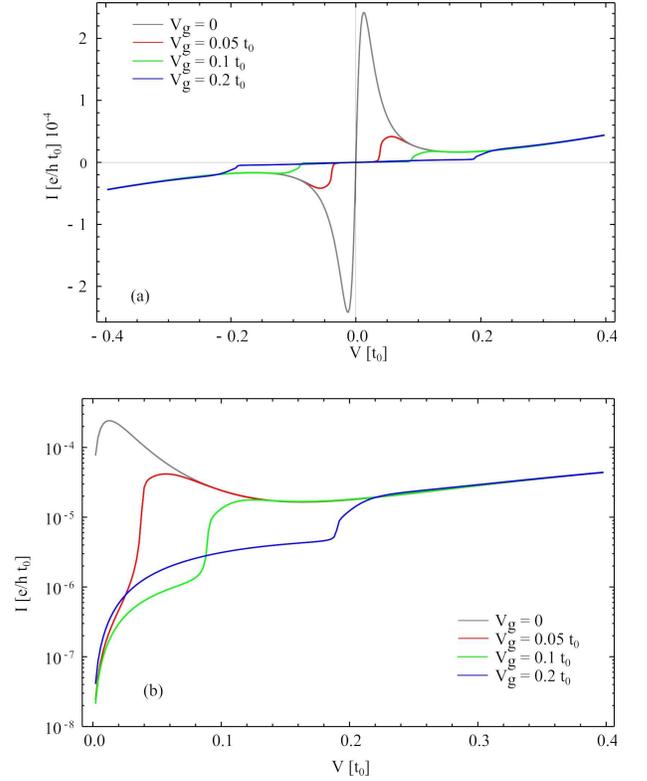}
\caption{I-V characteristics of graphene TFET at different gate voltage.}
\end{figure}
In order to explain the obtained specific I-V behavior let us consider the schematic electronic DOS diagram for edge atoms in Fig. 4.
\begin{figure}
\includegraphics[width=8cm]{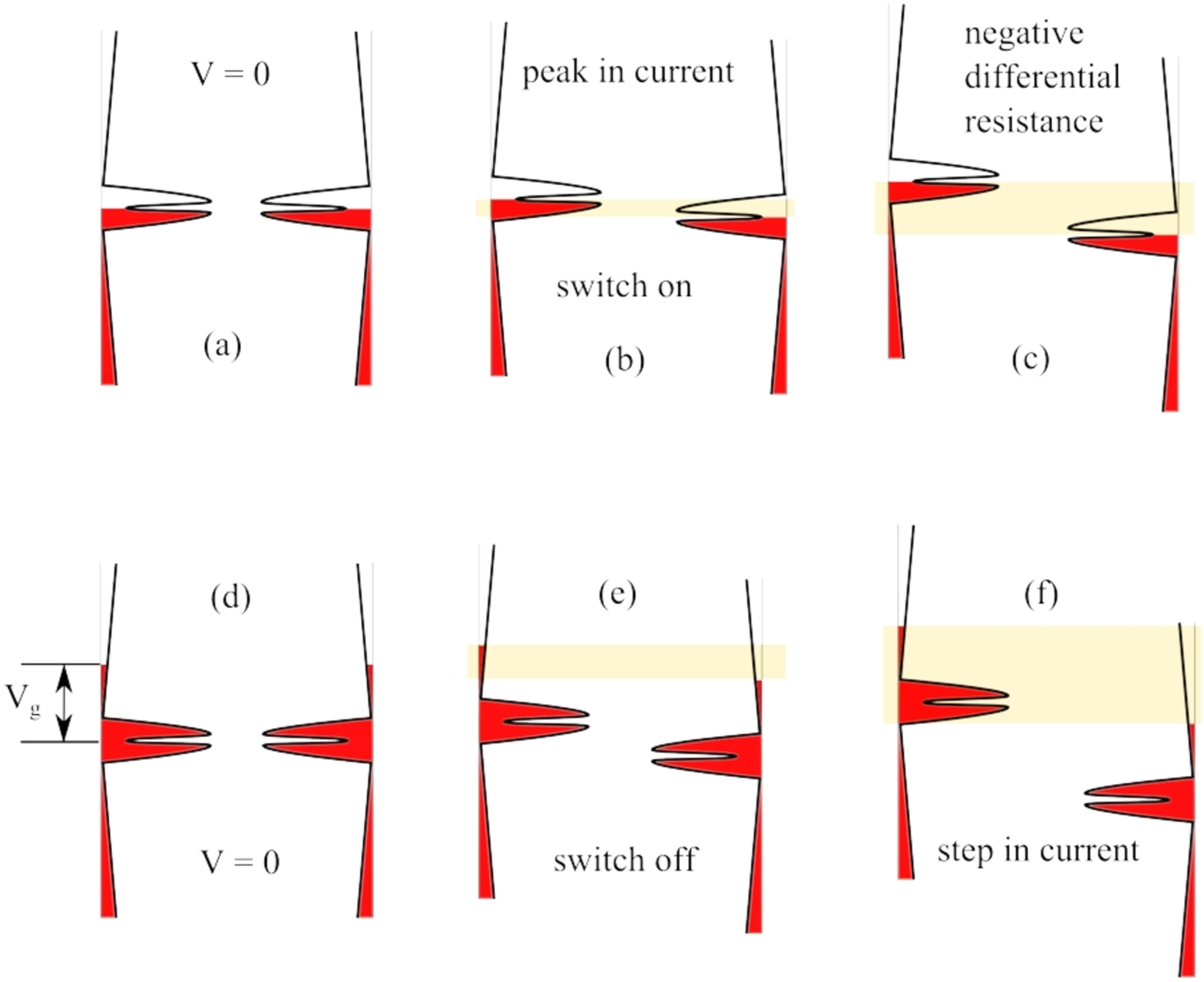}
\caption{Schematic electronic DOS diagram. The filled states have a red colour. Yellow region shows the energy window: carriers can now tunnel into empty states in the right contact.}
\end{figure}
Generally, in TFETs tunneling of interest is band-to-band tunneling. As was mentioned above, graphene electrodes terminated in the zigzag configuration have  pronounced peaks in the DOS at the Fermi level. At zero gate voltage the picture  is symmetric (Fig.4 (a)) and tunnel current is absent. Applied bias voltage provokes a shift of the bands and an opening of the energy window, which makes it possible for carriers to tunnel into empty states of the right electrode. This causes a short near to linear portion of the I-V curve at low bias voltage in Fig. 3. The maximum tunnel current takes place in the configuration shown in Fig. 4(b) when the filled states of the DOS peak in the left electrode are situated exactly opposite to the empty states of the DOS peak in the right electrode. Further increase in bias voltage leads to a larger shift of the peaks so that opposite states have a reduced DOS (see Fig. 4(c)). In this case, the tunnel current comes down and a region with negative differential resistance emerges in Fig. 3.  

Applied gate voltage will cause a shift of the Fermi level to the region of the low DOS in both contacts and, as a result, tunnel current turns out to be strongly suppressed (see Figs. 4(d),(e)) until one of the DOS peaks meets the energy window (Fig. 4 (f)). Fig. 1 (b) shows a possible schematic diagram of the I-V measurement setup. 
Two graphene layers are located on an insulator substrate opposite to each other.  Ohmic contacts are formed to both layers
individually representing the source (S) and the drain (D). The bottom gate voltages control the Fermi levels in the left and right layers of
graphene. Notice that a similar manipulation of the Fermi levels in graphene layers was proposed for Symmetric graphene TFET (SymFET)~\cite{SymFet}. 

\begin{figure}
\includegraphics[width=8cm]{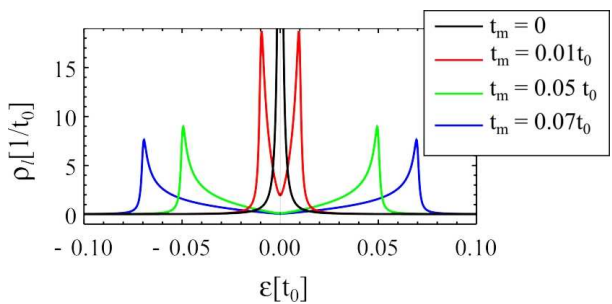}
\caption{Local DOS of edge atoms at different interaction constants between tunnel contacts.}
\end{figure}
In order to validate the above-presented picture we have calculated the local DOS of edge atoms in graphene contact. As is shown in Fig. 5, the local DOS depends on the interaction constant $t_m$ which, in turn, is related to the gap between electrodes.
For a large distance between contacts $t_m$ goes to zero and one obtains the sharp peak. When the gap decreases, the picture changes drastically: the DOS at the Fermi level decreases and there appear two peaks at the positions $\pm t_m$ around the Fermi energy.  In the limiting case of zero gap, $t_m=t_0$ and $\rho_l$ takes shape typical for an infinite graphene layer (see, e.g. Ref.~\onlinecite{NovoselovRev}).  Therefore, in order to get a pronounced switching effect one has to match the width of the gap between graphene electrodes taking into account that, on the one hand, the interaction parameter $t_m$ should be small enough and, on the other, the tunnel current decreases with  $t_m$ as $I\sim t^{2}_m$ (at $t_m \ll t_0$).
For example, at large enough $t_m$ the switching effect disappears as much as the peaks in the DOS do not fall into the energy window. In this case, the only influence of gate voltage is to vary slightly the DOS near the Fermi level, which will result in a modest increase of the conductance. Our analysis shows that such behavior takes place in tunnel contacts with armchair boundaries where the local DOS of edge atoms coincides with that for bulk atoms at the Fermi level~\cite{JLETT}. 

As is seen in Fig. 3, tunnel current depends on both gate and bias voltages. The on/off ($V_g = 0/V_g \neq 0$) current ratio reaches a value of the order of $10^3$ at $V < 2t_m$ and this quantity decreases with increasing $V$.  Raising gate voltage provokes a slight increase of the DOS near the Fermi level (see Fig. 4 (d)) thus increasing tunnel current and  decreasing the on/off ratio.  When $V$ exceeds $V_g$, the current becomes independent from gate voltage. 

An interesting question concerns the influence of the finite width of the contacts. Our analysis shows that the presented I-V curves are valid for the contact's width of the order of $10$ nm and more. For smaller widths the pictures are found to be similar, however, the curves have marked sub-peaks due to size quantization effects.  

In summary, we demonstrated a variant of planar graphene TFET which becomes switched-off at finite gate voltage. The current on/off ratio  
can reach a value of the order of $10^{3}$. The operation of the device is based on a possibility to manipulate the positions of peaks in the DOS of zigzag graphene edges with relation to the energy window. The good operation conditions  take place in a restricted region of bias voltage.  
We expect that the proposed mechanism for planar graphene TFET is quite reliable and can be revealed in experiment. To this end, the existing graphene nanogap fabrication methods should be improved in order to produce controlled nanometer-size gaps. 
Our device demonstrates an original switching mechanism and manifests an interesting region of the negative differential resistivity. There is reason to hope that all these findings may have extensive applications in graphene-based electronics.

\begin{acknowledgments}
This work has been supported by the Russian Foundation for Basic Research under grant No. 12-02-01081.
\end{acknowledgments}

\end{document}